\documentclass{aa} 

\newcommand{\kms}{\,$\mathrm{km\, s^{-1}}$}
\newcommand{\ms}{\,$\mathrm{m\, s^{-1}}$}

\usepackage{txfonts}
\usepackage{graphicx}
\usepackage[authoryear]{natbib}

\usepackage{longtable} 

\bibpunct{(}{)}{;}{a}{}{,} 

\begin{document}

\title{ Search for giant planets in M67  I. Overview 
\thanks{Based on observations collected at ESO, La Silla, Chile,  OHP and  HET.}
}
  

\author{L. Pasquini\inst{1} \and A. Brucalassi\inst{4,10}  \and M.T. Ruiz\inst{2} \and P. Bonifacio\inst{3} \and C. Lovis\inst{5} \and R. Saglia\inst{4,10} \and C. Melo \inst{7} \and K. Biazzo\inst{1,8} \and S. Randich\inst{7} \and L. R. Bedin\inst{9} }

\offprints{L. Pasquini, \email{lpasquin@eso.org}}

\institute{ESO -- European Southern Observatory, Karl-Schwarzschild-Strasse 2, 85748 Garching bei M\"unchen, Germany
 \and Universidad de Chile, Santiago, Chile  
\and  GEPI, Observatoire de Paris, CNRS, Univ. Paris Diderot, Place Jules Janssen 92190 Meudon, France 
 \and Max Planck F\"ur Extraterrestrische Physik, Garching bei M\"unchen, Germany 
 \and Observatoire de Geneve, Sauverny, CH
  \and ESO -- European Southern Observatory,Santiago, Chile
\and Istituto Nazionale di Astrofisica, Osservatorio Astrofisico di Arcetri, Firenze, Italy
\and Istituto Nazionale di Astrofisica, Osservatorio Astronomico di Capodimonte, Napoli, Italy
\and Istituto Nazionale di Astrofisica, Osservatorio Astronomico di Padova, Padova, Italy
\and University Observatory Munich, Ludwig Maximillian Universitaet, Scheinerstrasse 1, 81679 Munich, Germany
}

\date{Received  / Accepted }

\abstract
{ Precise stellar radial velocities are used to search for massive (Jupiter masses or higher) 
exoplanets around the stars of the open cluster M67. }
{ We aim to obtain a census of massive exoplanets in a cluster of solar metallicity and age in order 
to  study the dependence of planet formation on  stellar mass and  
 to  compare in detail the chemical composition of stars 
with and without planets. This first work  presents the sample and the observations, 
 discusses the cluster characteristics and the radial velocity (RV) distribution of the stars, 
and individuates the most likely planetary  host candidates. } 
{We observed a total of 88  main-sequence stars, subgiants,  and giants all highly probable 
members of M67, using four telescopes and instrument combinations. 
We investigate whether exoplanets are present 
by  obtaining radial velocities with precisions as good as  $\simeq$10 \ms. 
To date, we have performed 680 single observations (Dec. 2011) and a preliminary analysis of data, spanning a period of up to eight years. 
After reducing all the observations to the HARPS zero point, the RV measurements for each star are used to 
evaluate the RV variability along the cluster color magnitude diagram.} 
{Although the sample was pre-selected to avoid the inclusion of binaries, we identify 11 previously unknown binary candidates. 
The RV variance (including the observational error) for the bulk of stars is almost constant with 
stellar magnitude (therefore stellar gravity)  at $\sigma = $20 \ms. 
Eleven stars clearly displayed larger RV variability and these are 
candidates to host long-term substellar  companions. 
The average RV is also independent of the stellar magnitude and evolutionary status, confirming  
that the difference in gravitational redshift between giants and dwarfs is almost cancelled
by the atmospheric motions. We use the subsample of solar-type 
stars to derive a precise true RV for this cluster (M67$_{RV}$ = 33.74 $\pm$ 0.12 \kms) . 
The velocity dispersion is  0.54  \kms for giants and 0.68 \kms for dwarfs, substantially lower than reported in previous works. 
We finally create a catalog of binaries  and  use it to clean the color magnitude diagram (CMD).
Isochrone fitting confirms an age of around 4 Gyr. Further cleaning of the CMD based on precise RV 
could establish M67 as a real benchmark for stellar evolutionary models.  }
{ By pushing the search for planets to the faintest possible magnitudes, it is possible to observe solar analogues 
in open clusters, and we propose 11 candidates to host substellar companions.}

   
\titlerunning{Exoplanet  in M67 - I}
\authorrunning{L. Pasquini et al.}
\maketitle

\section{Introduction}
\label{sec:Intro}

\subsection{Searching planets in open clusters}
After the discovery of the first exoplanet around 51 Peg about 16 years ago  (Mayor and Queloz 1995), 
more than 700 planets have been discovered, but a number of very basic questions still await  answers, such as: 
how does the rate of planet formation depend on 
stellar metallicity and  mass? Does  planet formation strongly depend on  the stellar environment? 

Most exoplanets have been found around bright and nearby field stars that 
cover a very large range of stellar characteristics. These studies have  many advantages, but 
the widely differing characteristics  of 
field stars may also limit our capability to derive precise conclusions. 
For instance, it is puzzling that main-sequence stars hosting  giant planets 
are  metal rich (Gonzales et al. 1997, Santos et al. 2004), 
while  evolved stars that host giant planets are not (Pasquini et al. 2007). Is this because of 
stellar pollution acting on main-sequence stars (e.g. Laughlin \& Adams 1997), 
or because planet formation favors the  birth of planets around more metallic stars (Pollack et al. 1996) ?
Or maybe the metal-rich planet-hosting stars belong to an inner disk population, 
as proposed by Haywood (2009)? 
Are planet-hosting  stars more depleted in Li
than their non-hosting twins, as suggested by e.g. Israelian et al. (2009) (see however
Baumann et al.  2010 for a different opinion) ? Or are the volatile elements depleted in the convective zones
of stars that created rocky planets, as implied by the findings of Melendez et al. (2009) 
(see, however Gonzalez-Hernandez et al. 2010 for a different opinion)? 

A large number of planets discovered around stars in open clusters would provide the perfect sample to 
answer  all the above questions. Under the very reasonable assumption that open cluster stars have similar  
ages and 
chemical compositions (Pasquini et al. 2004, Randich et al. 2005, De Silva et al. 2007),  a thorough, 
detailed  chemical analysis of stars with and without planets belonging to the same cluster would provide 
 direct answers to all the questions and hypotheses above, without any need for further speculation. 

Similarly, it is now quite clear that stellar mass has a great influence on the 
frequency of giant planets (Lovis \& Mayor 2007, Johnson et al. 2010), although the precise dependence of the 
planet rate on stellar mass is not yet known. 
In addition, a large number of giant planets around stars of open clusters, 
which have precisely  determined masses, 
would provide an ideal database to study this dependence. 

Finally, although most stars were born in stellar clusters and star associations, very little is known 
about the frequency of planetary systems in different environments, and on how their survival and  evolution changes 
with it. A lot will be learned about this by directly imaging star forming regions, 
but the comparison between field and cluster statistics will help us to understand this point.

\subsection{ Previous searches}
In spite of the importance  of finding planets in open clusters, the literature on the subject 
is so far rather limited. 
The 
search for planets in open clusters with the RV technique has been  limited to the studies of Paulson et al. (2002, 2004) 
around main-sequence stars of the Hyades, and the search for planets around evolved stars in 
a few clusters by Lovis \& Mayor (2007) and around the Hyades giants (Sato et al. 2007). 
Paulson et al. did not find any evidence of short-period giant planets around the Hyades dwarfs, 
and excluded a high rate of hot jupiters in this high metallicity cluster. 
Several potential long-term candidates were present in their sample, but they  did not perform any 
follow-up for longer periods. 

Sato et al. (2007) found a long-period giant planet around one of the Hyades clump giants, and 
that there are only three such stars in this cluster led the authors to conclude  that stellar mass has a significant 
influence on the giant planet rate. 
Finally, Lovis \& Mayor (2007) collected  RV observations of evolved stars in open
clusters for several years. They found evidence of a couple of  sub-stellar mass objects, and used 
statistical arguments to conclude that a higher stellar mass is more likely to promote the formation of massive planets. 
 

Although the probability of finding a transiting exoplanet in an open cluster is rather low
 (van Saders \& Gaudi 2011), 
open clusters (OC) have been targeted for extensive transit
searches in the past few years(Bruntt et al.\ (2003), 
Street et al.\ (2003), 
von\ Braun et al.\ (2005),
Bramich et al.\ (2005), 
Mochejska et al.\ (2005), 
Burke et al.\ (2006),
Aigrain et al.\ (2006),   et al.\ (2007, 2011),
and Hartman et al.\ (2009, and references therein).  
However, only a handful of weak, unconfirmed candidates have been so far identified 
(Mochejska et al. 2006;  et al. 2011).

Being much richer in stars than OCs,  globular clusters (GCs) are of greater statistical
significance, particularly in  the case of a null detection (which always
seems to be the case).
Only a few GCs so far have been  systematically searched for exoplanet transits. 
Among these, we may  mention the ground-based campaigns targeting
47\,Tucanae (Weldrake et al. 2005) and  $\omega$\,Centauri (Weldrake et
al. 2008), both searching for hot-Jupiters around upper main-sequence
 stars in the outskirts of these two clusters, and both providing  
no significant exoplanet transit candidate.   

The systematic photometric search that had placed the tightest constraints 
on the planet-frequency occurrence in a cluster, was that of Gilliland et al.\ (2000). 
This work was based on \textit{Hubble Space Telescope} (HST) WFPC2 images of the 
dense core of 47\,Tuc, which was monitored for $\sim$8.3 days during which 
about 34,000 upper main-sequence  stars were observed. 
Gilliland et al.\ (2000) found no exoplanet transits, and  
concluded that the planet occurrence in 47 Tuc is smaller by a factor
of ten than for field stars. 

 Nascimbeni et al.\ (2012) analyzed a similar HST data
set (based this time on ACS/WFC images) in an outer field of the GC
NGC\,6397.   For the first time, this work focused on searching for 
hot-Jupiters among low mass stars (K and M spectral types).  Again,  
no high-significance planetary candidate was detected, but
owing to the lower quality statistics no firm conclusion was reached about the
occurrence of giant planets in  M and K stars of NGC\,6397.

The cause of the lack  of close-in planets in GCs is not fully understood. 
Presumably  the low metallicities and/or the dense environments  
interfere with planet formation, leading to  orbital evolution to close-in
positions, and/or  planet survival.

\subsection{ Why a search in M67?} 

M67 is one of the most well-studied open clusters.  It has been comprehensively observed to establish astrometric membership 
(Sanders 1977, Girard et al. 1989, Yadav et al. 2008), precise photometry 
(Montgomery et al. 1993, Sandquist 2004), and a  rather precise RV and binary 
search (Mathieu et al. 1986, Melo et al. 2001, Pasquini et al. 2011). 
X-ray sources have been identified (Pasquini \& Belloni 1998, Van den Berg et al. 2004),
 and was one of the 
first clusters for which observations of stellar oscillations were attempted (Gilliland et al. 1991). 
Its chemical composition and age are very close to  solar values (Randich et al. 2006, Pace et al. 2008, 
\"Onehag et al. 2010) and it hosts very good candidates for solar twins
 (Pasquini et al. 2008, \"Onehag et al. 2010) . 

For an open cluster, M67  is quite rich in stars,  and its color magnitude diagram (CMD) 
is well populated in  
the main sequence, in the subgiant and red giant  (RGB) branches. 
With a distance modulus of 9.63 (Pasquini et al. 2008) 
and a low reddening (E(B-V)=0.041, Taylor 2007), the solar stars have an apparent  magnitude of V=14.58 
and a (B-V) of 0.69 (Pasquini et al. 2008), and the cluster contains more than 100 stars brighter than this magnitude 
suitable for a RV planet search. 

\section{Sample and Observations}
\label{sec:Obs}
We selected stars that are   proper motion members with a probability higher than 60$\%$ from 
Yadav et al. (2008) and also RV members, and unknown  binaries 
from previous studies. 
In this context we recall the extended work  of Mathieu and coworkers (Mathieu et al. 1986),
who made a very complete RV survey of the evolved stars of M67 with a precision of a few hundred \ms. 

The majority of the other stars were selected following  Pasquini et al. (2008), who
used several VLT-FLAMES exposures for each star to classify suspected binaries. 
The full sample includes a total of 88 stars, from solar type (faintest is V$\sim$15) 
to the tip of the RGB. 
The stars are rather faint for precise radial velocity observations, but a RV  precision of 
$\sim$10\ms can be obtained for each  measurement even for the faintest objects, with  observations 
shorter than one hour at the ESO 3.6m telescope. 

The bulk of the observations were carried out with HARPS at the ESO 3.6m telescope 
(Mayor et al. 2003), and this instrument 
is our reference for all the observations. 
Given the superior performances of HARPS, we concentrated mostly on the faintest  objects with this facility. 

After the project started, we  added the  sample of evolved stars observed by CORALIE in the years 
2003-2005, as part of a program of planet search of giants in open clusters  (cfr. Lovis \& Mayor 2007) .
One limitation of our program is the sparse sampling frequency of the observations: 
typically a few nights/yr  were 
awarded in the period January-April, and large gaps, longer than six months, 
are present between one season of observations and the next. 
We tried to gather HARPS observations for stars all over the CMD, 
 to have enough points to derive proper zero-point 
offsets for the other instruments. 
Figure  1 presents the CMD of M67, using the photometry of Yadav et al. (2008) and stars with at least 
60 $\%$ membership from proper motions. The sample stars are marked in green,  and
spectroscopic binaries (see Table \ref{table:Binaries}) in red.

\begin{figure}
\includegraphics[width=9cm]{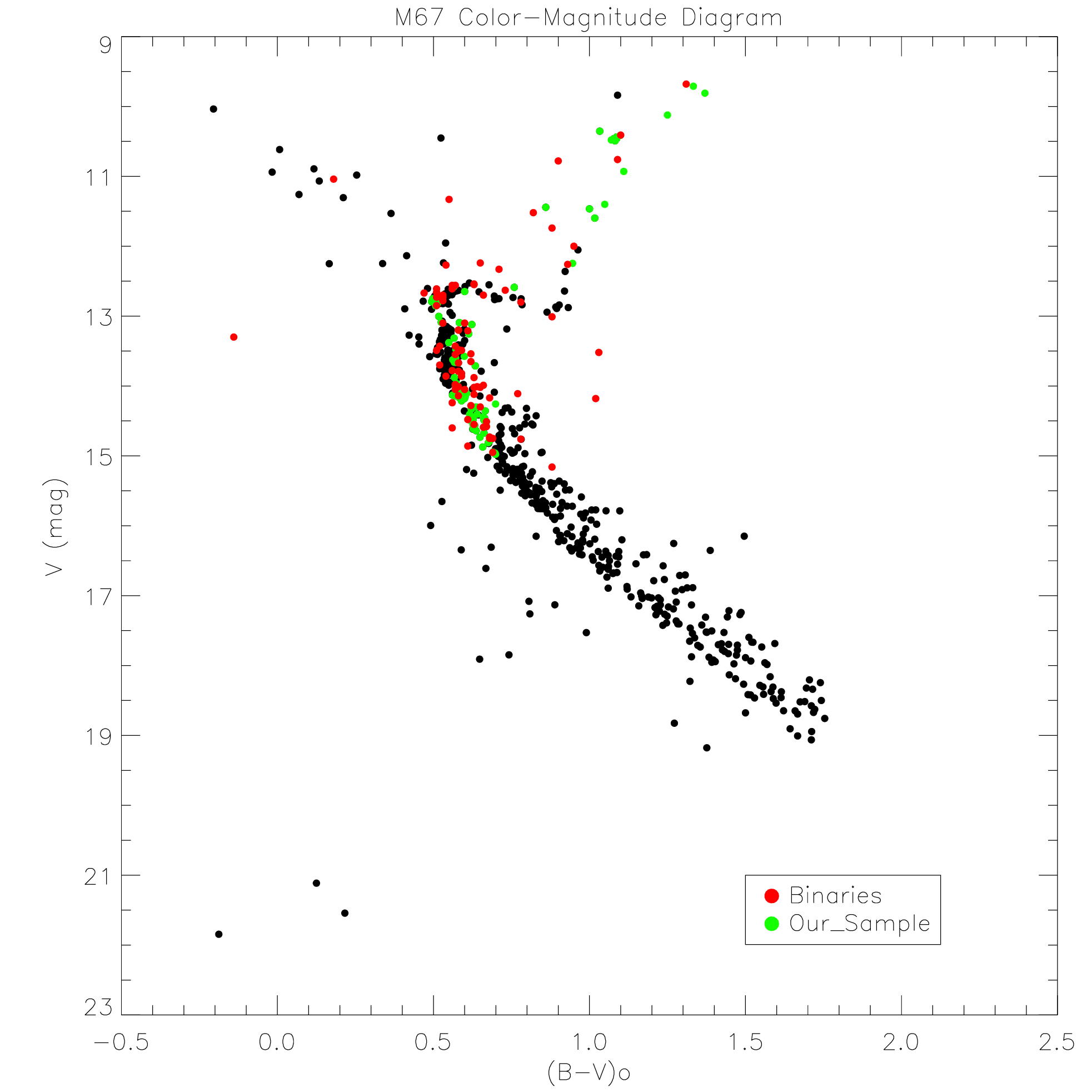}
\caption{ Color-magnitude diagram (CMD) of M67. The photometry is from Yadav et al. 2008. 
Only stars with a high membership  probability 
($\ge$60 $\%$) are shown. Known binaries, either from this work or the literature (cfr. Table \ref{table:Binaries}) 
are shown in red colors. The stars  observed in this survey are marked in green.   } 
\label{fig:cmd}
\end{figure}

\subsection{HARPS observations}
HARPS (Mayor et al. 2003) is the planet hunter at the ESO 3.6m telescope. 
In high accuracy mode (HAM) it has an aperture on the sky of one arcsecond, 
and a resolving power of  115000. The  spectral range covered is 380-680 nm. 
In addition to be exceptionally stable, 
HARPS achieves  the highest precision  using the simultaneous calibration principle:  
 the spectrum of a calibration (Th-Ar) source is recorded simultaneously  with the stellar spectrum,
 with  a second optical fibre. 
Since the M67 stars are quite faint for this instrument, we opted to use HARPS in the  high efficiency mode: 
the fibre has a larger aperture on the sky (1.2 arcseconds, corresponding to 
R=90000) and is not equipped with an optical scrambler. 
This mode is limited to a precision of a few (5-7) \ms, but it is 30-40 $\%$ more efficient than the HAM mode. 
For our purposes this precision is sufficient, and the improved efficiency ensures that a high enough signal-to-noise (S/N) ratio 
can be obtained even for the faintest stars. 
As a rule of thumb we can consider that the precision of HARPS scales as $\epsilon_{RV} \propto 1/(S/N)$ (see below). 
Since the aim of  this giant planet survey is a precision of the 
single measurement of $\sim$ 10 \ms, it is possible to reach our goal with limited S/N  observations, of on
the order of S/N=10 at the peak of the signal. 
As  a consequence,  we can limit  the 
integration time to less than one hour even for the faintest stars. 
Our HARPS spectra have typically a peak S/N  of 15 for the faintest stars. 

HARPS is equipped with a very powerful pipeline that provides on-line  RV measurements, 
which are computed by cross correlating the stellar spectrum with 
a numerical template mask. This on-line pipeline also
provides  an associated RV error. For all of our stars, irrespective of the 
spectral type and luminosity, we used the solar template (G2V) mask. 

\begin{figure}
\centering
\includegraphics[width=9cm]{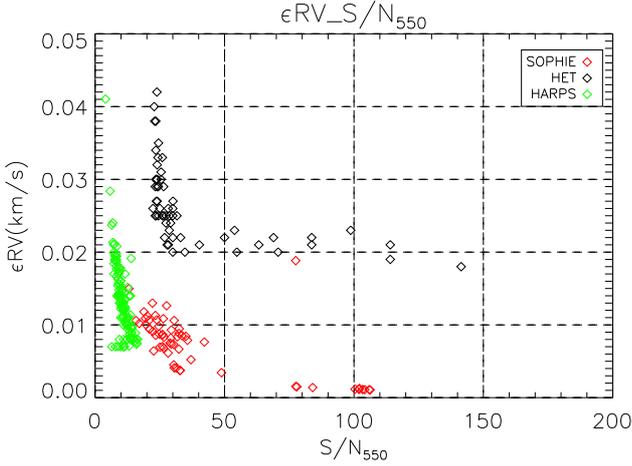}
  \caption{Errors in the RV measurement of single observations 
  vs. S/N  at 550 nm for HARPS (green points, only observations of MS faint stars), 
 SOPHIE (red points, all measurements, typically TO and evolved stars) and HET 
(black points, all measurements,mostly TO and evolved stars) are presented. 
The precision of the majority of the  measurements for the faint stars is 
between 10 and 20 \ms. 
For HARPS, systematic dominates the uncertainty at 8\ms for S/N's above 
$\sim$13 in EGGS mode. For our SOPHIE observations 
we estimate that systematic effects (not included in the figure) dominate the error below $\sim$12 \ms. 
The V mag range of the star sample for SOPHIE and HET is brighter than HARPS. } 
  \label{stono}
 \end{figure}

Figure \ref{stono} shows the error associated with the HARPS RV measurements versus (vs.) 
the S/N  of the observations computed at the middle of echelle order 50 (555 nm) for the 
faintest stars of the sample (Vmag $>$14). The RV precision 
scales approximately as $\epsilon_{RV} \sim 100/(S/N)$ when expressed in \ms and it levels off, 
as expected,  at 8 \ms for S/N above 13 for this order. 
 In the figure, the uncertainty associated with each RV measurement is also  
given for SOPHIE (red) and HET (black). 
The magnitude range of the star sample is between 10.0-14.5 for SOPHIE and 9.8-14.0 for HET.

Between January 2008 and March 2011 we  gathered 
409 observations of 88 stars with HARPS, which represent 
the bulk of this work. 

\subsection{SOPHIE observations} 

SOPHIE is the planet hunter at the 1.93m OHP telescope (Bouchy et al. 2006). 
The instrument concept and data reduction is similar to that of HARPS; 
in high efficiency mode  it has an aperture on the sky 
of three arcseconds with which a resolution of 40000  
is obtained.  In our first observing runs we observed  all M67 stars with SOPHIE, including solar twins, 
but the smaller telescope
diameter  and the somewhat more critical weather  conditions in winter at OHP than at  La Silla,
prompted our decision  not to use this instrument further for the faintest objects. 

We considered 34 M67 stars in common between our targets observed with SOPHIE and HARPS 
to compute the zero point between the two instruments. 
The comparison gives  RV (SOPHIE)= RV HARPS -12.34$\pm$8.0 \ms, 
with no dependence on the spectral type. 
This value is  confirmed by the observations 
of the star 104Tau (HD32923), for which a difference of -11.40$\pm$7.0 \ms between the two
 instruments is found (for a sample of 11 observations with HARPS and 5 with SOPHIE). 
We finally adopt a zero-point offset of 
-11.40 \ms  between SOPHIE and HARPS. 

We analyze 78 SOPHIE observations of M67 stars with an associated precision of $\simeq$12 \ms. 
Since the observations with HARPS and  the 
other instruments were not simultaneously acquired, we can assume that the 
precision with which the offsets are computed also includes the contribution 
from the intrinsic variability of the stars (and of course the photon and the instrumental noise).

\subsection{CORALIE observations}
CORALIE is located at the 1.2m Euler Swiss telescope at La Silla (Baranne et al. 1996). 
The M67 stars were observed between 2003 and 2005 in the framework of  a 
larger program of search for planets around giants in open clusters (Lovis \& Mayor 2007). 
The technique used to measure the RV in CORALIE observations is again the same as described for HARPS. 
As for the other instruments, the zero-point shift to HARPS was computed by using observations 
of stars in common to both instruments.
Since the stars in common are only giants, it is expected that the intrinsic RV 
variability of these stars is larger than for main-sequence 
objects (Setiawan et al. 2004), and not negligible.  
We used ten stars in common between HARPS and CORALIE to evaluate the offset, obtaining 
RV(Coralie)= RV HARPS + 26.8 $\pm$ 5.0 \ms. We have so far gathered 123 observations for 17 giants
with CORALIE with a precision associated with these observations of $\simeq$20 \ms. 

\subsection{HET Observations}

HRS, mounted  on the 10 m HET telescope (Tull 1998),  was the last instrument used in our  survey. 
We were granted 70 observing runs in service mode between November 2010-April 2011.
Each run consisted of two exposures of 1320 s and counts as one observation.
The configuration was set to a wavelength range between 407.6 nm and 787.5 nm with a central 
wavelength at 593.6 nm and a resolving power of R=60000. 
We were able to observe 13 objects selected from our sample with 9.0$\le$Vmag$\le$14.6.
The S/N  for the faintest stars is $\sim$ 10.
The radial velocities were computed using a series  dedicated  routines (Cappetta et al. 2012,  in preparation). 
The different steps include the wavelength calibration using a Th-Ar lamp exposure performed before and after each 
stellar spectrum, the normalization of the spectra, 
the cleaning of cosmic rays and both, telluric and sky lines, 
the computation of the heliocentric corrections and finally the 
cross-correlation of the spectrum with a G2 star template.
We used the multiple exposures of the same star to estimate the typical error bar associated 
with the HET observations, finding an error of $\simeq$25 \ms.   

When considering the different analysis used for the HET data with respect to the other instruments, 
it is unsurprising to find a larger offset  with respect to HARPS for HRS than for the other instruments: 
RV (HRS)= RV(HARPS)+ 242.0 $\pm$ 12\ms ( eight stars were used for the comparison). 
{\bf 
\begin{table}
\label{teltintable}
\begin{tabular}{l|cccc}
  Istrument & \multicolumn{1}{c}{HARPS} & \multicolumn{1}{c}{SOPHIE} & \multicolumn{1}{c}{HET} & \multicolumn{1}{c}{CORALIE } \\ 
\hline
N obs.Stars  &  88          &     54        &      15       &     17        \\
Observations &  409         &     78        &      70       &    123        \\
MS stars     &  58          &     42        &       7       &               \\
TO stars     &   7          &      2        &       4       &               \\
G stars      &  23          &     10        &       4       &     17        \\
Period       &  2008-2011   &  2008-2011    &  2010-2011    &   2003-2005   \\

\end{tabular}
\caption{Table presenting the number of observed stars, the total number
         of observations, the number of main-sequence (MS), turn-off (TO),
         giant stars (G) observed for each instrument.}
 \end{table}
}

\begin{figure}
\centering
\includegraphics[width=9cm]{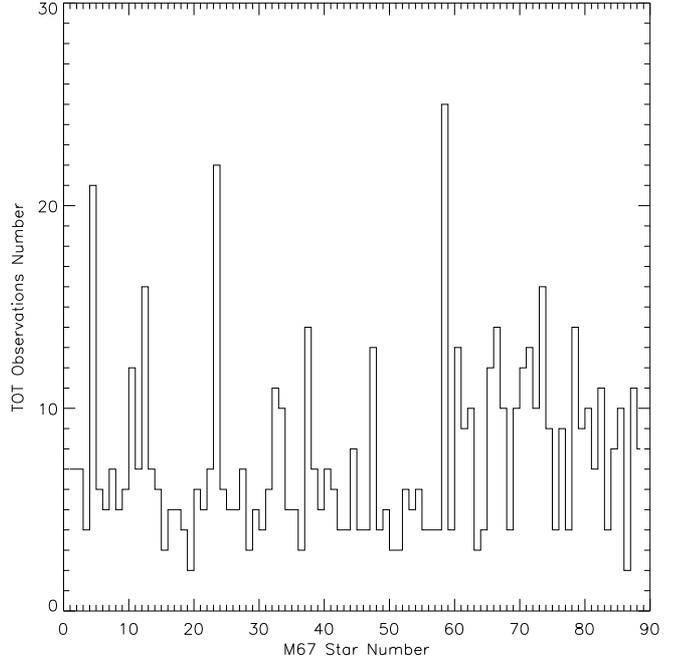}
\caption{Histogram showing the  number of observations/star for our total sample.
 All observations from HARPS, SOPHIE, CORALIE, and HRS are included in the plot.  We have observed each star on average seven times.  }
\label{N_Obs}
\end{figure}

\section{ Results}
Once corrected to the zero points of HARPS, all the observations of each star were
collected and analyzed together. 
Table \ref{table:observations} summarizes the main data for the observed stars. 
In addition to the basic stellar parameters, the number of 
observations per star is given, for each spectrograph and as a total. 

We have obtained, on  average, 7 observations/star, but this  
ranges from a minimum of 2  to more than 20 (see Table \ref{table:observations}). The histogram showing the 
number of observations/star is given in Figure \ref{N_Obs}. We performed a series of Monte Carlo 
simulations to establish the minimum number of observations/star that we need to exclude
 the presence of hot Jupiters at high confidence.  The preliminary results indicate that a final
 number of at least 9 observations/star should be reached (Brucalassi et al. 2012, in preparation). 

In the last two columns of Table\ref{table:observations} the mean stellar RV of each star is given,  
together with the RV dispersion. 
Individual RV measurements will be provided in a forthcoming paper devoted to the detailed 
discussion of the planets' host candidates and to discussion of the presence of hot Jupiters
 (Brucalassi et al. 2012, in preparation).
 

\subsection{Binaries and CMD}

\begin{figure*}
\centering
\includegraphics[width=10cm]{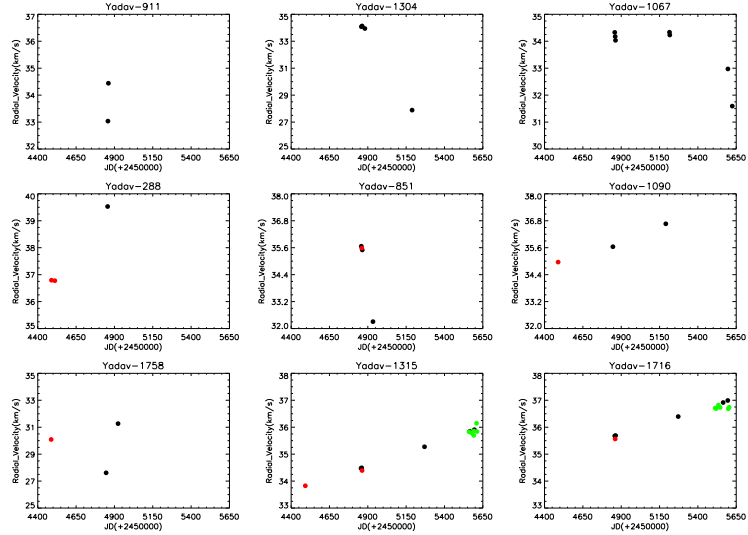}
 \caption{Radial velocity curves of 9 binary candidates.
 The minimum span in RV is about 1.7 \kms. The error bars are not shown. Different colors refer to HARPS (black), SOPHIE (red), and
HET (green) observations. The other two binary stars not shown have double-line CCF, so are double-line spectroscopic binaries. } 
  \label{radvelbin}
 \end{figure*}

One of the first findings of our survey is that, despite  all the stars having been previously 
observed and found to have no evidence of companions,  11  stars
in the original sample of 88 (13$\%$) show RV variations that are too large to be produced by  
an exoplanet, or by a non-stellar object. 
We considered as binary candidates all the stars displaying a peak-to-peak
RV amplitude of at least 1.7 \kms. Considering half of the difference as a lower limit to the 
orbital semi-amplitude,   this amplitude 
corresponds to  a companion of 15 Jupiter masses on a 30 day period for a circular orbit around a star of 
1.2 solar masses. 
The RV range spanned by these stars is so large that planetary companions can be 
excluded, as can be seen from Figure \ref{radvelbin},
where the RV measurements for 9
of the binary candidates are shown. The binary/long-term RV variable nature of 7 of them was  confirmed by D. Latham (private communication), 
who is performing a  long-term RV monitoring of  more than 400 M67 stars (Latham 2006). 

\begin{table}
\caption{The binary candidates of our  sample}
\label{radvelbintable}
\begin{tabular}{lcccc}
  Object & \multicolumn{1}{c}{V} & \multicolumn{1}{c}{B-V} & \multicolumn{1}{c}{RV$(km/s)$} & \multicolumn{1}{c}{$\sigma$RV$(km/s)$ } \\ 
\hline

  Y$288$  & $13.9$ & $0.637$ & $37.691$ & $1.299 $ \\
  Y$769$  & $13.5$ & $0.641$ & $CCF double-peaked$ &  \\
  Y$851$ &  $14.1$ & $0.617$ & $34.759$ & $1.417    $ \\
  Y$911$ &  $14.6$ & $0.673$ & $33.738$ & $ 0.703   $ \\
  Y$1090$ & $13.8$ & $0.650$ & $35.186$ & $ 1.265  $ \\ 
  Y$1304$ & $14.7$ & $0.723$ & $32.512$ & $ 2.670   $ \\
  Y$1758$ & $13.2$ & $0.653$ & $29.653$ & $ 1.521   $ \\
  Y$1315$ & $14.3$ & $0.693$ & $34.885$ & $0.801$ \\
  Y$1716$ & $13.3$ & $0.619$ & $36.205$ & $0.651$ \\
  Y$1067 $ &$14.6$ & $0.642$ & $33.667$ & $1.030$ \\
  S$1583$ & $B=13.1 $ &      & $CCF double-peaked$ &  \\ 
 \hline
 \end{tabular}
 \end{table}

The measurements are also  given in table \ref{radvelbintable},  which is available in electronic form. 
These stars are binary candidates, and 
were not observed  after  a large variation of their RV  was measured. 
Given that these stars are high-probability M67 proper-motion members, and that their RV is close 
to that of the cluster,  it is very 
likely that they are spectroscopic binaries belonging
to the cluster.

Two stars (S815 and S1197, cf Table \ref{table:observations}) show peak to peak RV variations of 
the order of 700 /ms ;  they are retained in the single star sample, although the amplitude of the RV variation is 
possibly too high to host a planet. 

In the process of evaluating the effects of binaries in M67 CMD, we found that 
 binaries are identified in many works in 
the literature  and sometimes different names are used; in addition  
several works have been published after the compilation of Sandquist (2004).
We  therefore opted to create a new catalog of binaries in M67  that includes our candidates, binaries from the literature, as well as binary candidates from  X-ray observations.  
The catalog is given in Table \ref{table:Binaries}, with reference to the original studies.  

The binary stars from Table\ref{table:Binaries}   are plotted in Figure 1 as red points. 
The high percentage of binaries in M67 is unsurprising, given that 
to retain the stars for such a long time, M67 has the most massive stars in the core, and  
some mass segregation  has occurred.

\begin{figure}
\centering
\includegraphics[width=9cm]{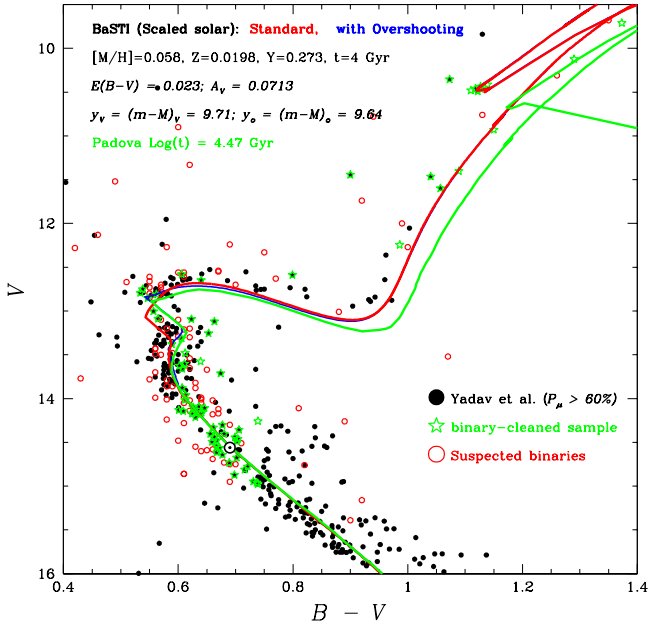}
  \caption{ Color-magnitude diagram (CMD) of M67 (photometry from Yadav et al. 2008) for
probable members ($P_{\mu}>60\%$, filled dots).
Probable single stars within our sample are indicated with star symbols, and probable binaries with empty
dots. The location of the Sun, as if it were within M67, is marked with the $\odot$.
The 4-Gyr isochrones in red and blue are from the {\sf BaSTI} library (Pietrinferni et al. 2004) for standard models and with overshooting, respectively. 
The isochrone in green is a 4.47 Gyr from Girardi et al. (2000). } 
  \label{mod}
 \end{figure}

A complete  census of the  binaries in M67 is helpful because this cluster 
can be used to test the effects of several mechanisms  debated in stellar evolution, such as 
diffusion  and overshooting (see e.g. the discussion in Magic et al. 2010).  
Cleaning of the CMD is especially important in the region of the turnoff, 
because, as is clear in
Figure \ref{mod},  in that region binaries cannot be photometrically distinguished from 
single main-sequence stars. 
They separate  more clearly along the main sequence, where a 
separate, detached binary sequence is present, although  several fainter binaries lie on 
the main sequence and are photometrically indistinguishable from single stars in the CMD.

In his extensive study of M67, Sandquist (2004) created  one table containing the 
fiducial sample of single stars and a second table 
with a list of interesting or peculiar stars. 
In the first list, 11 stars are indicated as possible binaries according 
to our catalog (S1305, 1458, 990, 1300, 1075, 1201, 982, 1102, 1452, 951, 820), 
and one star (S1197) has large RV variations. 
Among these 11 binary candidates, some  have  weak evidence coming  
from a few FEROS or FLAMES multiple spectra, which had a limited precision. For instance star  S1305, 
a low-RGB star, is indicated as a suspected binary in Pasquini et al. 2011, 
but is not confirmed by our higher precision measurements (cfr. Table\ref{table:observations}). 
Other Sandquist' single-star sequence fiducial stars (e.g. S982, 1201, 1452) are, on the other hand,  
confirmed to be binaries by our high-precision RV measurements. 

Three stars of the  Sandquist 'unusual stars' table are  confirmed to be RV 
multiple candidates 
according to Table \ref{table:Binaries} (S1292, S816, S1011). 

The above results show that, in spite of the large efforts to clean the  CMD of M67, a number of 
unknown binaries are still present and  the detailed comparisons 
required to distinguish  between different potential mechanisms (Magic et al. 2010) 
could strongly benefit from additional  cleaning, in particular around the turnoff.

In Figure \ref{mod}, we show the observed region of the CMD with  the isochrones
from  Pietrinferni et al.( 2004), with and without overshooting.
In this CMD, we also indicate the position of the solar analog, 
as determined in Pasquini et al. 2008, and we impose that the tracks appropriately fit this point, 
in addition to the rest of the CMD. The  4 Gyr track with moderate overshooting seems to most closely
represent the data.
In the same figure, we also superimpose the Padova isochrones,  with 
solar metallicity, age 4.47Gyr, and Y=0.26 (Girardi et al. 2000, as from Girardi web page). 
The isochrone fits  the turnoff very well, but  produces a 
RGB and clump that are too red. The mismatch is not dramatic and may indicate   
some problem in either the bolometric correction used or  some of  the  free  parameters adopted 
(e.g. mixing length). 
We note that for both sets of isochrones a slightly lower reddening 
(E(B-V)=0.02 instead of 0.041 (Taylor 2007)) is needed to match the colors of the turnoff. 
We also note  that there are a number of stars, 
apparently with constant RVs, and a high probability of proper motion membership, that 
are above the main sequence.

This analysis confirms that, in addition to its extraordinarily similar abundance pattern, 
M67 has  an age compatible with that of the  Sun. 

Given these similarities and that the solar birthplace has not yet been identified, it 
is natural to ask  whether M67 and the Sun  were associated in the past. This problem was
exhaustively studied by Pichardo et al.(2012), who performed a full set of 
dynamical simulations,  
excluding that M67 and the Sun were born in the same cloud. This result leaves us therefore  with a
question mark about the birthplace of the Sun and might challenge the validity  
of chemical tagging, which associates stars  with either clusters or 
associations  based on the  similarity of their chemical compositions
 (see e.g. Freeman and Bland-Hawthorn 2002).

\subsection{ Radial velocity variability }

We  first  investigate  the observed  RV variability along the CMD. 
Since all the stars belong to the cluster with high probability, we have  the unique opportunity to study how 
RV variability changes along the CMD diagram, for a given chemical composition and age.
In addition,  a diagram with RV variability vs. magnitude  would immediately highlight any
 possible 'outliers', which would  be the most obvious  hosts of  exoplanets.

Setiawan et al. (2004) and Hekker and Melendez (2007) shwed that the intrinsic RV variability of giants
increases with stellar luminosity and becomes large for bright, low-gravity giants. 
Since we cover a six magnitude interval, it could be useful to determine for each 
magnitude or evolutionary status a typical average RV variability. This 
quantity should depend solely on both, the intrinsic stellar RV variability and 
the RV measurement error  (typical photon errors associated with the bright stars are smaller  
because faint star observations are limited by photon noise).

\begin{figure}
\includegraphics[width=9cm]{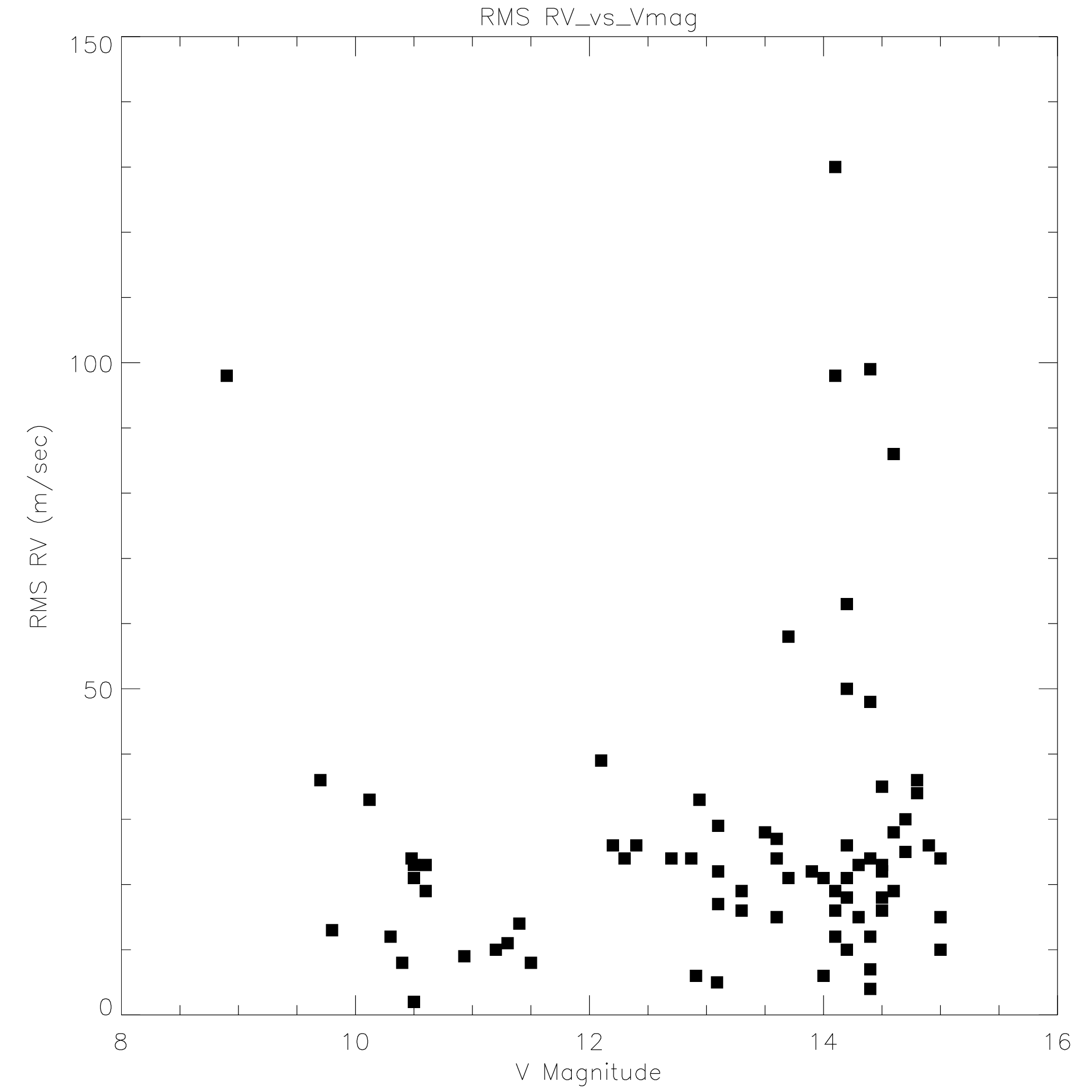}
\caption{Radial velocity (RV) variability  of the observed stars vs. V magnitude. The bulk of the 
stars show a flat behavior. This is likely a combination of larger intrinsic variability for the 
more luminous stars and larger measurement uncertainty  for the faintest ones. The excess of variability 
seems real, and may indicate that more stars have low mass companions.}
\label{rmsRVmagV}
\end{figure}

Figure \ref{rmsRVmagV} shows the rms RV  vs. V diagram for 75 single stars. 
Binary candidates have been excluded, and  the two stars (S815 and S1197) 
with high RV variability are not shown. 

Some of the stars with very little or no RV variability have very few observational points, and their 
small scatter is very likely the result of our low data statistics.  
The increase in RV scatter with stellar luminosity observed in field stars (Setiawan et al. 2004) 
is not  evident in our sample. There is a possible hint of an increase in the range 
of magnitudes 12$<$V$<$14, but 
the RV variability does not increase further for the more luminous stars.  

As a general conclusion, we can say that the RV variability shown in  Figure \ref{rmsRVmagV}
is basically constant at 20 \ms ($\sigma$), independent of the stellar magnitude. 
The bulk of our observations have a RV variability that is well represented by a Gaussian distribution 
centered at 20 \ms of width $\sigma$ = 10 \ms.

The  unexpected flatness of the RV variability with magnitude is most likely due to the combination of two 
effects. In evolved stars, some measurable  stellar RV variability is present, while for  the faint 
main-sequence stars the uncertainty in the measurements increases because of the limited S/N. 

\begin{figure}
\includegraphics[width=11cm]{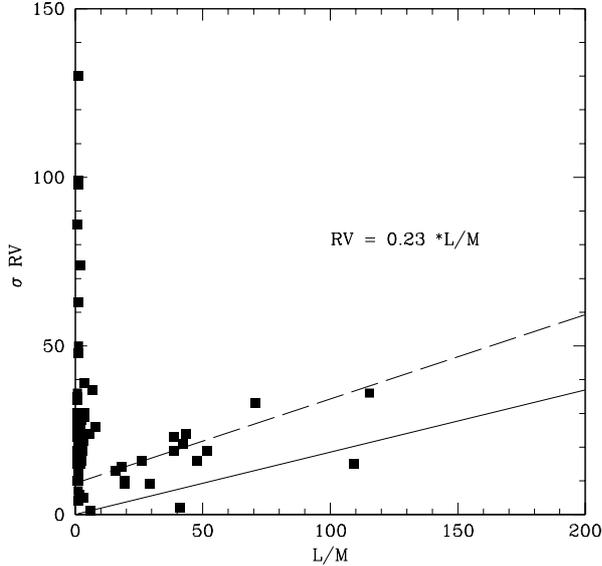}
\caption{Radial velocity variability $\sigma$ RV  of the observed stars vs. the luminosity/mass ratio. 
The solid line represents the scaling law proposed by Kjeldsen and Bedding (1995) to extrapolate solar type 
oscillations to other stars. To plot the scaling law a fixed value (1.2 M$_{\odot}$) for the mass 
has been assumed. To first approximation all M67 evolved stars have the same mass. The continuous line 
is the scaling law, the dashed line is the best fit to the evolved stars data, which is perfectly 
consistent with the scaling law, when measurement uncertainties are taken into consideration.}
\label{RVrmsL/Mratio}
\end{figure}

To investigate these points, in Figure \ref{RVrmsL/Mratio} 
we plot the RV variability as a function of the  
stellar luminosity to mass ratio (L/M). 
According to Kjelsden and  Bedding (1995) the RV  jitter induced by 
solar oscillations is expected to grow according to the law RV $\propto$ 0.23*L/M (with 
RV expressed in \ms). This law is represented by the continuous line. 
Figure \ref{RVrmsL/Mratio}  illustrates the very good agreement between this scaling law 
and the RV variability of evolved stars in M67. The continuous line remains just below the measurements, 
but this does not take into account  that the measurements, in addition to the stellar intrinsic 
variability, include measurement uncertainties that are not negligible. 
A fit to the evolved star data (L/M $>$ 5)  gives $\sigma_{RV}$ = 0.25 L/M + 9.3 (\ms),  which is 
 shown by the dashed line in Figure \ref{RVrmsL/Mratio}.   
The observations and the predictions therefore match quite well, when the measurement  errors are
considered.  We note that this comparison implies that the oscillation RV amplitude and its variations are of 
comparable size. 
More precise   RV measurements would allow us to quantitatively investigate this point in more  detail. 
We conclude that the M67 evolved stars show signs of intrinsic RV variability and that 
for most of the evolved stars the observed  RV variability is consistent with that 
expected from the Kjelsen and Bedding (1995) scaling 
law for solar-type  oscillations.

\begin{figure}
\includegraphics[width=9cm]{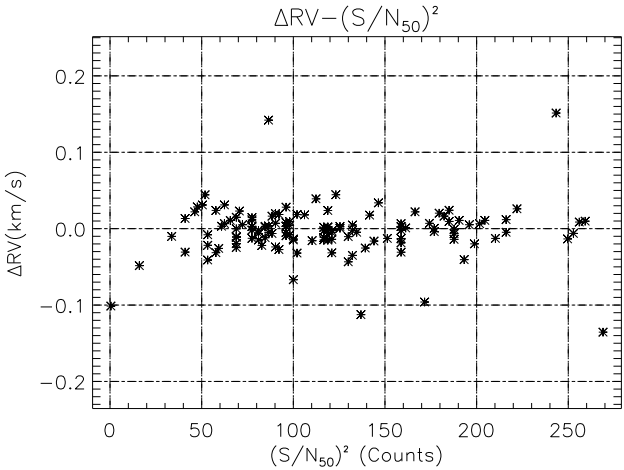}
\caption{Difference in RV measurement from the average for faint stars in the sample vs. 
the square of the S/N  at 550 nm, a quantity that is roughly proportional to the 
flux recorded in the spectra. No trend is visible, and this indicates that if
detector charge transfer inefficiency is present, it does not affect our measurements in a detectable way.  }
\label{rvflux}
\end{figure}

As far as the lower main sequence is concerned, 
the behavior of the  faint stars in 
our sample (which are also the least massive ones) is quite interesting, because if the RV variability 
observed were induced by substellar companions, this would  already indicate a correlation 
between planet frequency and mass in M67.  
There are good arguments to believe that 
 an RV variability as large as 20 \ms cannot be 
 caused by intrinsic stellar noise, because stellar
 noise for solar stars scales with activity and therefore age, and M67 is almost as old as the Sun  
(see e.g. Saar et al. 1998, Dumusque et al. 2011).
On the other hand we have seen  that the uncertainties in the RV measurements of faint stars is 
above 10 \ms (cfr Figure \ref{stono}) but well below 20 \ms. 
 We  investigated whether other instrumental effects, 
not included in the data analysis, could affect the RV measurement precision at low count levels.  
At least two effects could influence the observations, and we  investigated 
whether our measurements depend on observational parameters, such as the observed flux or 
airmass. A dependence on flux could be induced, for instance, by CCD transfer inefficiency, 
which has been  reported  to be high in SOPHIE (Bouchy et al. 2009). 
A dependence of RV on airmass 
 could instead indicate that some systematic effects are induced  by the HARPS atmospheric 
 dispersion compensator  at high airmass or by the guiding system of the telescope. 
For every solar star, we therefore computed  the $\Delta$RV of each observation with respect to the average 
stellar radial velocity, and analyzed all the measurements of all stars together, 
as functions of counts and airmass. No trend is present, as is clear from Figure \ref{rvflux}, 
where the $\Delta$RV is plotted vs. the square of the S/N  at 550 nm. 
We conclude therefore that these two quantities do not affect our measurements in an appreciable way. 
We note that in these comparisons only the  HARPS data have been used, because they 
by far dominate the faint star statistics.

With a RV variability for the whole sample  centered at 20 \ms and  a 
width  $\sigma$ of 10 \ms, we can safely  assume that stars with a
RV variability at or above 50 \ms are very good candidates for low mass companion hosts. 
A number of stars (9) stand out clearly from the general constant trend of Figure 3, showing a 
$\sigma$ RV variability of 50 \ms or larger. 
These stars (plus S815, and S1197, which  are not included in the figure) are candidates to host
 giant planets 
or substellar objects. 
Another 6 candidates have smaller, but still interestingly large RV variability,  
and deserve to  be investigated further. 
The list of the most likely 11  candidates is given in Table \ref{cdsan1}. 

\begin{table}
\caption{Stars with the largest RV variability, which are candidates to host substellar companions}
\label{cdsan1}
\begin{tabular}{lcccc}
  Object & \multicolumn{1}{c}{V} & \multicolumn{1}{c}{B-V} & \multicolumn{1}{c}{RV$(km/s)$} & \multicolumn{1}{c}{$\sigma$RV$(km/s)$ } \\ 
\hline

  Y$401$  & $13.7$ & $0.566$ & $33.203$ & $0.058$ \\
  Y$673$  & $14.4$ & $0.665$ & $33.766$ & $0.099$ \\
  Y$1051$ & $14.1$ & $0.595$ & $33.290$ & $0.130$ \\
  Y$1587$ & $14.2$ & $0.600$ & $33.434$ & $0.063$ \\
  Y$1722$ & $14.2$ & $0.560$ & $34.460$ & $0.098$ \\ 
  Y$1788$ & $14.4$ & $0.622$ & $34.150$ & $0.048$ \\
  Y$1955$ & $14.2$ & $0.589$ & $33.192$ & $0.050$ \\
  Y$2018$ & $14.6$ & $0.631$ & $31.953$ & $0.086$ \\
  S$488$ & $8.9$  & $1.550$ & $32.910$ & $0.089$ \\ 
  S$815$    &  $12.9$   &  $ 0.497$ &  $33.326$   &   $ 0.378$    \\
\hline
\end{tabular}
\end{table}

Several stars show 
clear long trends of RV variability compatible with the presence of a planet, but the 
baseline is not yet long enough to determine the nature of the companion. More RV points
 are required and  are being acquired. A full analysis of the data of our completed  
survey will be published elsewhere (Brucalassi et al. 2012). 

The large scatter in the radial velocities, in excess of the measurement errors,  
may indicate that more stars, in addition to those indicated in Table \ref{cdsan1} 
are suitable candidates for hosting planets.

Keeping in mind that we are only sensitive  to rather massive planets, we find it
interesting that our  candidates should all have long orbital periods. 
Mayor et al. (2011) found clear evidence that the most massive planets tend to have  
 long orbits, which is perfectly in line with our results. 
Similarly, if all 9 candidates were planets, they would correspond to  a frequency of 
  giant planets of $\sim$ 13$\%$, which  agrees with the rate of giant planets found by 
 Mayor et al. (2011) and  by   D\"ollinger et al. (2012, in preparation) around evolved stars. 

\subsection{ Gravitational redshift in M67 and  cluster radial velocity}

 Pasquini et al. (2011)  used FEROS and literature observations of M67 stars to investigate 
whether gravitational redshift could be detected in the stars of this cluster, by comparing the  
measured RV of the cluster dwarfs and giants. 
Their adopted technique  assumes that the subsamples of cluster stars share the same cluster RV,
 irrespective of their mass and evolutionary status.

These authors found that, when using the radial velocities derived from cross correlation masks, 
there was no evidence of gravitational reddening: giants and dwarfs have, within the RV 
uncertainties, the same radial velocity. 
They also showed that this behavior is  compatible with the shifts and asymmetries of spectral lines 
 predicted by 3D models. They found in addition that  M67 giants are dynamically cooler than dwarfs. 

Although our sample is smaller than the one used by Pasquini et al. (2011), our RV data are 10-20 times more 
precise, and, in addition, the many observations acquired have allowed us to eliminate  several binaries, 
which would cause the cluster to appear dynamically hotter and perhaps introduce 
a skewness in  the RV distribution. 
Since all our newly discovered  binaries   are on the main sequence, 
they might have influenced the conclusions about the dynamical status of the cluster, making the dwarfs 
appear dynamically hotter. 
Pasquini et al. (2011) derived a $\sigma_{giants}$ = 680 \ms and a $\sigma_{dwarfs}$= 900 \ms 
and estimated that their FEROS RV precision is of $\sim$300 \ms. The RV precision of our measurements is more than 
one order of magnitude more accurate than these values. For the first time we 
have gathered a sample of stars in M67 whose  RV measurement errors are definitely 
negligible with respect to the cluster dynamics. 

\begin{figure}
\centering
\includegraphics[width=10cm]{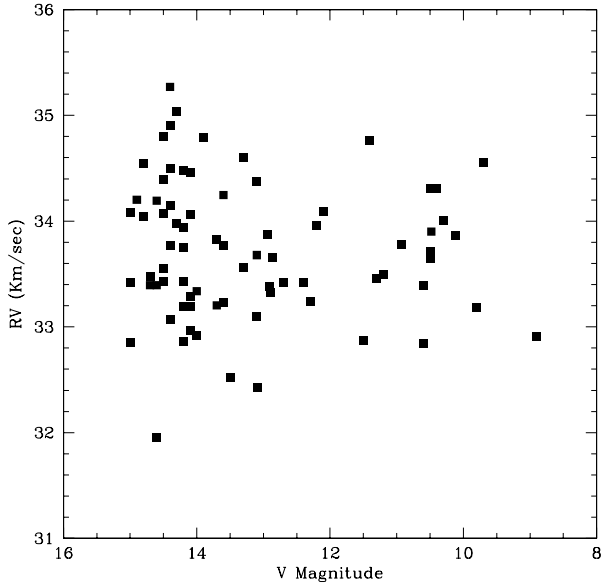}
\caption{M67 stellar magnitude - RV diagram for our observations. No dependence of the stellar RV 
on the stellar magnitude is present. The data confirm previous findings that the dwarfs are dynamically hotter than 
giants.  }
\label{magrv}
\end{figure}

\begin{figure}
\centering
\includegraphics[width=10cm]{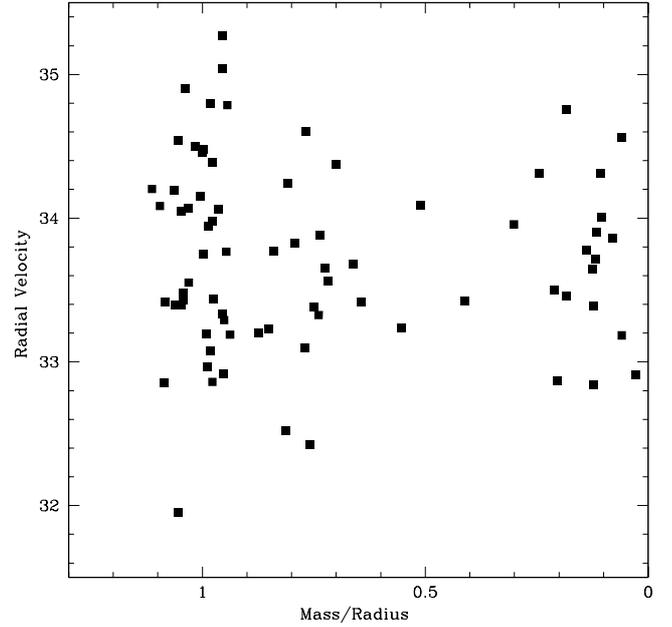}
\caption{M67 radial velocity vs. M/R for all  stars. No strong dependence of the stellar RV 
on  M/R  is present. Gravitational redshift would predict a 0.6 slope. }
\label{msur}
\end{figure}

Figure \ref{magrv} shows the distribution of  stellar RV vs.  stellar magnitude,  
and Figure \ref{msur} the same RV but vs. the ratio of  stellar mass to radius (see Pasquini et al. 2011 for details).
 If gravitational reddening were the only process acting, we would expect a 
dependence of  RV = RV$_0$ + 0.632(M/R).

There is clearly  no  significant dependence of RV  on either   magnitude or  M/R, 
confirming the FEROS results 
of Pasquini et al. 2011. A least squares fit  to the V-RV diagram gives 
$ RV = 33.362 + 0.0295*V$, where the angular coefficient  is closely compatible with zero (error  is 0.047).

The average radial velocity of all stars is 33.724 kms$^{-1}$, with a dispersion of  $\pm$0.646 \kms. 
By dividing the sample into ``giants'' and ``dwarfs'' 
at B-V=0.7, we find $\sigma_{giants}$ = 540 \ms ($\pm$ 90 \ms, 18 stars, v=33.67) and  
$\sigma_{dwarfs}$ = 680 \ms ($\pm 63$ \ms, 59 stars, v=33.74). 

By computing the ratio mass to radius  for each star (cf. Pasquini et al. 2011) we find 
that RV depends only very slightly on M/R, with an angular coefficient of  0.096 that is much 
smaller than the 0.6 expected from gravitational reddening (cf. Figure 6), 
confirming  the results of Pasquini et al. (2011). 

In the future we plan to merge all the  HARPS spectra acquired for each star to perform 
a detailed spectroscopic study,  including chemical abundance and line-shift analyses.

Since the present RV measurements are much more precise than in the past,  
the RV dispersion is smaller than that found by 
Pasquini et al. (2011). If we attribute the  
difference between the giant dispersion found in Pasquini et al. (2011) (680 \ms)   and that 
found in this paper (540 \ms)  entirely to the FEROS RV uncertainty, $\epsilon RV_{Feros} \sim
\sqrt( 680^2 -540^2) $, this would be approximately 400 \ms. The value is   
slightly larger than the FEROS errors estimated by Pasquini et al. (2011) (300 \ms), but  
 it is not unreasonable when considering that 
for most FEROS stars only one observation was available,  so some intrinsic stellar RV 
variability should also contribute to the difference. 

Finally we use the solar-stars radial velocities to compute the true radial velocity of the cluster.
We consider that, what is commonly called 'radial velocity' is 
the measure of the center of the cross-correlation function (CCF) between the stellar lines and a digital mask that in 
our case is based on the solar spectrum. 
This implies that the measured Doppler velocity  includes the stellar radial velocity, the effects of 
stellar atmospheres, and 
the errors in the positioning of the lines in the digital mask (see Lidegren and Dravins 2003).

Our observations span the whole cluster and include many solar-type stars. 
This  allows us to properly convert the measured Doppler shifts of the single stars 
into a measurement of the true radial velocity of the cluster.
 We  therefore analyze  only the 36 stars most similar to the Sun, in the range 14.0$<$ V $<$15. 
The magnitude of the closest solar analogs is 
V$\sim$14.58 (Pasquini et al. 2008, \"Ohenaag 2010). The only two  underlying hypotheses are that as a group 
these stars share the same cluster velocity and that they are representative of the solar spectrum. 
The average Doppler shifts for the CCF of these 36 stars is 
 M67$_{Ds}$=33.83 kms$^{-1}$, with a $\sigma$= 712 \ms, which implies that the associated 
uncertainty for M67$_{Ds}$ is  118 \ms.

To compute the true radial velocity of the cluster, we must correct in addition  for the 
zero-point offset of the G2 mask used. We did this by using the asteroid observations 
of Ceres obtained by HARPS and described in Molaro et al. (2008) and Molaro \& Centurion (2011). 
The expected radial velocity of the asteroids with respect to La Silla can be 
computed rather accurately, and the difference between the computed  and 
the measured radial velocity gives the 
zero offset of the mask from the solar spectrum. With this procedure, we can pass from 
Doppler shift measurements to the true radial velocity of solar stars. 

Table \ref{tablerv} shows the results of the expected vs. measured Ceres radial velocity: the two results are 
very consistent, to the order of  1 \ms, indicating a shift for the mask of 94.5 \ms. This velocity must be 
subtracted from the measured M67$_{Ds}$  to obtain the true cluster radial velocity, which is therefore
M67$_{RV}$ = 33.74 $\pm 0.12$ kms$^{-1}$.

\begin{table}
\caption{Predicted and measured CERES RVs. Mid Julian dates and predicted RVs are taken from Molaro et al. 
2008. In the last column, $\Delta$RV indicates RV measured - expected, and provides the zero-point offset of the HARPS
mask for solar stars, in \ms .  }
\label{tablerv}
\begin{tabular}{lccc}
  JD & Expected RV & Measured RV &  $\Delta$V \\ 
 Mid-exposure & kms$^{-1}$ & kms$^{-1}$ & \ms\\
\hline
  2453932.837256  & -10.832 & -10.738  & 94 \\
  2453877.919808  & -22.017 & -21.922  & 95 \\
\hline
\end{tabular}
\end{table}

We note that the error given by the zero-point correction of the mask is 
completely negligible compared to the 
uncertainty introduced by the cluster internal motions. Therefore, by enlarging the 
sample of solar-type stars, the final error could easily be diminished and very precise 
measurements of the radial velocity of the cluster 
could be  obtained. 
This mask offset can also be applied to solar-type field stars to derive their true radial velocities, 
and this method could be extended to determine precise  
RV offsets as a function of stellar gravity and temperature 
for solar metallicity stars. Finally,  true radial velocities could then be determined for 
field stars of different spectral types. 

Our new data confirm that the dwarfs are dynamically hotter (ratio 1.25)
than the giants in this cluster.  
All the derived parameters  agree very well with those computed by the maximum likelihood 
estimation in Pasquini et al. (2011). 

\section{Conclusions}

We have presented a long-term search program for giant planets in the solar-age, 
solar-metallicity open cluster M67.

We have used four different instruments, and, after finding proper zero-point corrections to HARPS, 
we have analyzed 680 observations for 88 stars. 
Twelve new binaries have been identified and we have created a catalog of known binaries in M67, 
which  we have used
to clean the cluster CMD. We have found that the CMD is well represented by tracks of solar age with some
overshooting.  The observed RV variability  does not depend very strongly on the 
stellar magnitude. 
The evolved stars  show a RV variability that follows the Kjeldsen and Bedding (1995) 
scaling law for solar oscillations quite closely, while the RV variability for the 
main-sequence stars is dominated by the RV measurement uncertainties. 

Eleven  stars show long-term  RV variability in a range that  
make them interesting candidates for exoplanet hosts. If confirmed, their long period and fraction would  
agree with recently derived  statistics for field stars (Mayor et al. 2011). 

We finally used our precise RV measurements to confirm that no gravitational redshift  is  measured 
between M67 giants and dwarfs, confirming  that the velocity dispersion of main-sequence stars is 
larger than that of 
giants (680 vs. 540 \ms,  respectively). We determined the zero-point shift of the G2 mask of HARPS 
and  by using the solar stars  we determined a true M67 radial velocity of 
33.74$\pm$0.12 kms$^{-1}$. 

We have shown that the search for planets in open clusters is a really powerful tool for investigating 
a large number 
of questions related to planet formation and stellar evolution. The new generation of spectrographs,
such as ESPRESSO (Pasquini et al. 2009), will make this search more effective and precise. 


\begin{acknowledgements}
LPA thanks ESO DGDF and the kind hospitality of UFRN(Br); he also thanks  Paul Bristow for a careful reading of the manuscript.

RPS acknowledges the support of RoPACS during this research, a Marie Curie Initial Training Network funded by the 
European Commissions Seventh Framework Programme. The Hobby Eberly Telescope (HET) is  a joint 
project of the University of Texas at Austin, the Pennsylvania State University, Stanford University, Ludwig Maximilians Universit\"at Muenchen, and Georg-August Universit\"at Goettingen. 

We are grateful to Programme National de Physique Stellaire
and Programme National de Planetologie of the Institut de Science
de l'Univers of CNRS for allocating and supporting the SOPHIE
observations.

\end{acknowledgements}




\setcounter{table}{4}
\begin{table*}
\caption{Observed targets in M67. Stars $B-V$ colors, apparent V magnitudes, and spectroscopic data. 
For object identifications, 'S' are from Sanders (1977); 'YBP' from Yadav et al.\ (2008). The full table is available in electronic form only.  } 
\label{table:observations}
\begin{tabular}{cccccccccc}
\hline
Object	&      $B-V$ 	   &        V    & Obs HARPS & Obs SOPHIE & Obs CORALIE & Obs HET & TOT & RV(\kms) & Sigma(\kms) \\
\hline
YBP266  &      0.570       &      13.6       &    5  &  2  &     &     &   7  &     33.773   &    0.027 \\
YBP285  &      0.663       &      14.5       &    5  &  2  &     &     &   7  &     34.391   &    0.018  \\
YBP291  &      0.570       &      13.5       &   18  &  3  &     &     &  21  &     32.524   &    0.028 \\
YBP349  &      0.636       &      14.3       &    6  &     &     &     &   6  &     35.040   &    0.023  \\
YBP350  &      0.561       &      13.6       &    4  &  1  &     &     &   5  &     33.228   &    0.024  \\
YBP401  &      0.566       &      13.7       &    4  &  3  &     &     &   7  &     33.203   &    0.058  \\
YBP473  &      0.658       &      14.4       &    5  &     &     &     &   5  &     35.269   &    0.024  \\
\end{tabular}
\end{table*}
\begin{table*}
\caption{Binary candidates in M67. Stars $B-V$ colors, apparent V magnitudes. 
For object identifications, 'S' are from Sanders (1977);'YBP' from Yadav et al.\ (2008). The full table is available in electronic form only.} 
 \label{table:Binaries}
\begin{tabular}{cccc}
 \hline
 Object	&      $B-V$ 	   &        V    &    Reference \\
 \hline
S251      & 0.67   &  12.55   &  IV,V,IV,VI      \\
S440      & 2.05   &  08.15   &  IV              \\
S1000     & 0.39   &  12.80   &  V,IV	     \\
S1011     & 0.63   &  13.82   &  I,VII           \\
S1040     & 0.49   &  11.52   &  V,IV,VI,VIII,IX      \\  
S1072     & 0.62   &  11.33   &  V,IV,IX	      \\ 
S1182     & 0.99   &  12.00   &  IV,VI	      \\
          &        &          &        	       \\


\multicolumn{4}{l}{References to original papers:}\\
\multicolumn{4}{l}{I:  Pasquini et al. (2011) }\\
\multicolumn{4}{l}{II:  Pasquini  et al. (2008)} \\
\multicolumn{4}{l}{III: Pasquini et al. (1997)} \\
\multicolumn{4}{l}{IV: Mathieu  et al. (1990)}  \\
\multicolumn{4}{l}{V:  Latham  et al.(1992)} \\
\multicolumn{4}{l}{VI: Mathieu  et al.(1986)}  \\
\multicolumn{4}{l}{VII: New binary candidates from our Sample} \\
\multicolumn{4}{l}{VIII: Pasquini \& Belloni (1998)} \\
\multicolumn{4}{l}{IX: Van den Berg  et al. (2004) }\\
\multicolumn{4}{l}{X:  Sandquist \& Shetrone (2003)}  \\
\multicolumn{4}{l}{XI:  Belloni, Verbunt \& Mathieu (1998)}  \\
\multicolumn{4}{l}{XII:  Sandquist et al. (2003)}\\ 
\end{tabular}
\end{table*}

\end{document}